\documentclass[sigconf,11pt]{acmart}

\usepackage{booktabs} 
\usepackage{caption}
\usepackage{subcaption}
\setcopyright{none}
\setcopyright{rightsretained}
\acmConference[FAT/ML]{4th Workshop on Fairness, Accountability, and Transparency in Machine Learning}{2017}{Nova Scotia, Canada} 
\acmYear{2017}
\copyrightyear{2017}

\acmBadgeR{}

\usepackage{flushend}

\usepackage[boxruled]{algorithm2e}
\usepackage{algorithmic}

\begin{document}

\title{Fair Personalization}
\subtitle{[Extended Abstract]}

\author{\large L. Elisa Celis}
  \authornote{Supported in part by SNF Project Grant 205121\_163385. }
\affiliation{%
  \institution{\normalsize{\'Ecole Polytechnique F\'ed\'erale de Lausanne (EPFL)}}
}
\email{elisa.celis@epfl.ch}

\author{\large Nisheeth K. Vishnoi}
\affiliation{%
  \institution{\normalsize{\'Ecole Polytechnique F\'ed\'erale de Lausanne (EPFL)}}
}
\email{nisheeth.vishnoi@epfl.ch}

\begin{abstract}
Personalization is pervasive in the online space as, when combined with learning, it leads to higher {efficiency and} revenue by allowing the most relevant content to be served to each user. 
However, recent studies suggest that such personalization can propagate societal or systemic biases, which has led to calls for regulatory mechanisms and algorithms to combat inequality. 
Here we propose a rigorous algorithmic framework that allows for the possibility to control  {biased} or discriminatory  {personalization} with respect to sensitive attributes of users without losing all of the benefits of personalization.
\end{abstract}

\maketitle

\newcommand{\parag}[1]{ {\bf \noindent #1}}
\newcommand{\defeq}{\stackrel{\textup{def}}{=}}
\newcommand{\nfrac}{\nicefrac}
\newcommand{\opt}{\mathrm{opt}}
\newcommand{\tO}{\widetilde{O}}
\newcommand{\polylog}{\mathop{\mbox{polylog}}}
\newcommand{\supp}{\mathrm{supp}}
\newcommand{\rank}{\mathrm{rank}}

\newcommand{\conv}{\mathrm{conv}}
\newcommand{\dist}{\mathrm{dist}}
\newcommand{\argmin}{\operatornamewithlimits{n}}
\newcommand{\sgn}{\mathrm{sgn}}

\newcommand{\cM}{\mathcal{M}}
\newcommand{\cB}{\mathcal{B}}
\newcommand{\cU}{\mathcal{U}}
\newcommand{\cP}{\mathcal{P}}
\newcommand{\cC}{\mathcal{C}}
\newcommand{\cY}{\mathcal{Y}}
\newcommand{\capa}{\mathrm{Cap}}
\newcommand{\dcapa}{\underline{\mathrm{Cap}}}
\newcommand{\st}{\mathrm{s.t.}}
\newcommand{\un}{\mathrm{un}}

\newcommand{\Pb}{\mathbb{P}}
\newcommand{\sym}{\mathrm{sym}}
\newcommand{\pcount}{\mathbf{PCount}}
\newcommand{\mixdet}{\mathbf{MixDisc}}
\newcommand{\sbold}{\mathbf{S}}
\newcommand{\mb}{{M(\cB)}}
\newcommand{\mlb}{{M_{\mathrm{lin}}(\cB)}}
\newcommand{\redc}[1]{ \textcolor{red} {#1}}

\newcommand{\eps}{\varepsilon}

\def\showauthornotes{0} 
\def\showkeys{0} 
\def\showdraftbox{0}

\newcommand{\Snote}{\Authornote{S}}
\newcommand{\Scomment}{\Authorcomment{S}}

\definecolor{Mygray}{gray}{0.8}
\newtheorem{remark}[theorem]{Remark}
\newcommand{\todo}[1]{\colorbox{Mygray}{\color{red}#1}}
\renewcommand{\C}{\mathcal C}

\newcommand{\nsimplex}{\Delta^n}
\newcommand{\dotprod}[2]{\left\langle #1, #2 \right\rangle}
\newcommand{\lnorm}[2]{\|#1\|_{#2}}
\newcommand{\vareps}{\varepsilon}
\newcommand{\pot}{\Phi}
\newcommand{\bestm}{m_{\star}}
\newcommand{\bestmT}{\bestm^T}
\newcommand{\bb}[1]{\left( #1 \right)}

\section{Introduction}

\begin{figure}[t]
\begin{center}
\includegraphics[width=\linewidth]{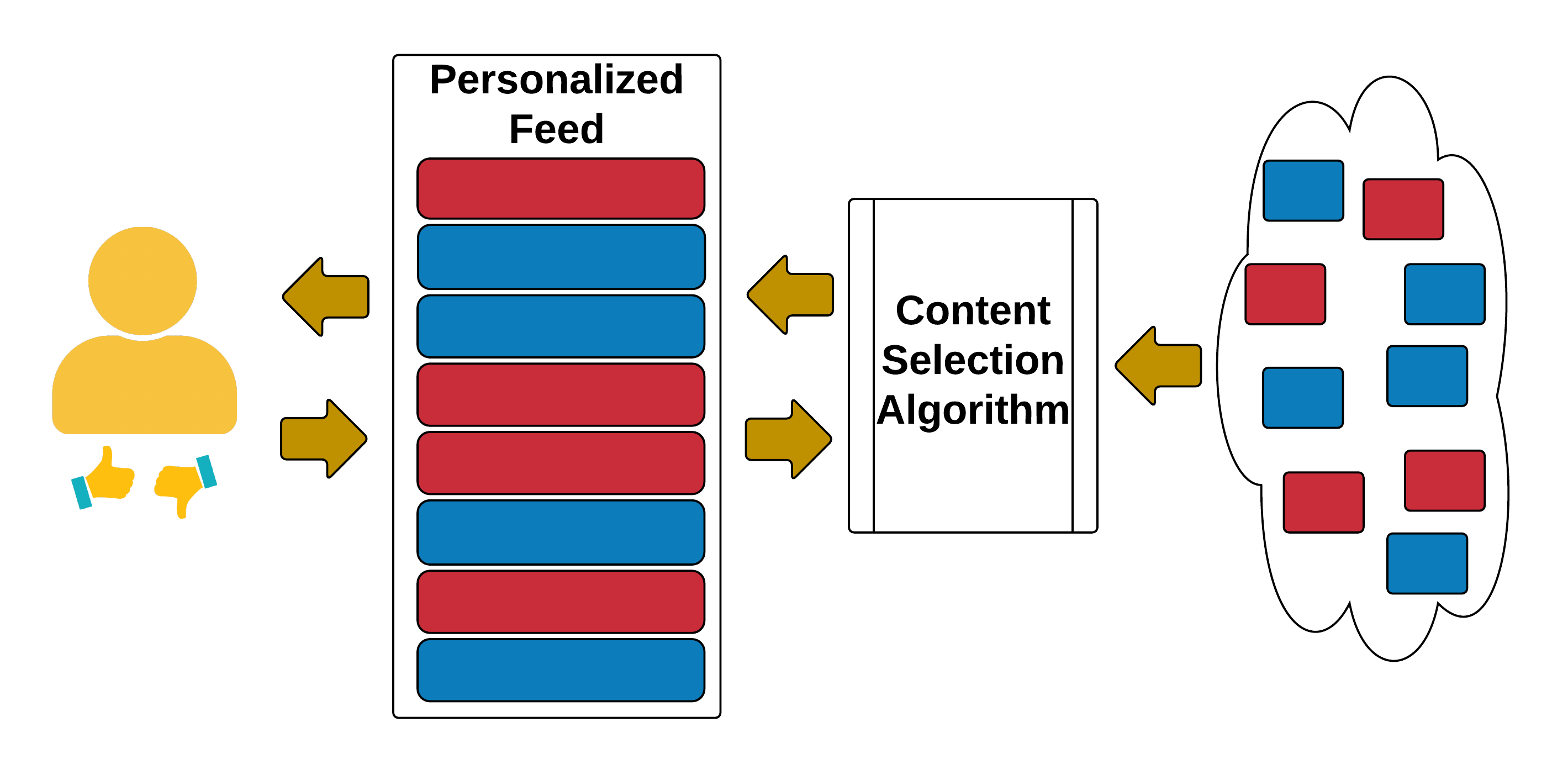}
\caption{{The content selection algorithm decides what to show to the user based on their properties and the feedback received from similar users. Different colors represent different types of content, e.g., news stories that lean republican vs. democrat. {Properties could be the gender, browser, location or friends of the user. Feedback could be past likes, purchases or follows.}}}
\label{fig:setting}
\vspace{-.2in}
\end{center}
\end{figure}

News and social media feeds, product recommendation, online advertising and other media that pervades the internet is increasingly personalized. 
Content selection algorithms take data and other information as input, and -- given a user's properties and past behavior -- produce a personalized list of content to display  \cite{jeh2003scaling,goldfarb2011online,liu2010personalized,carmel2009personalized} (see Figure~\ref{fig:setting}).
This personalization leads to higher utility and efficiency both for the platform, which can increase revenue by selling targeted advertisements, and also for the user, who sees content more directly related to their interests \cite{Forbes2017,farahat2012effective}.

However, it is now known that such personalization may result in propagating or even creating biases that can influence decisions and opinions. Recently, field studies have shown that user opinions about political candidates can be manipulated by biasing rankings of search results \cite{Epstein2015}. 
Similar concerns appear in other settings; e.g., it has been shown that ranking and classification algorithms can discriminate in online search results \cite{kulshrestha2017quantifying,FatML} 
and allow for gender inequality in serving targeted advertising for high-paying jobs \cite{datta2015automated, sweeney2013discrimination,farahat2012effective}.
As over two-thirds of American adults consume news online on social media sites such as Facebook and Twitter \cite{Mitchell2015}, and users are more likely to trust higher ranked search results \cite{Pan2007}, the potential impact of changing how social media ranks content is immense.

\begin{figure*}
\begin{center}
	\begin{subfigure}[b]{0.45\textwidth}
	\centering
		\includegraphics[width=\textwidth]{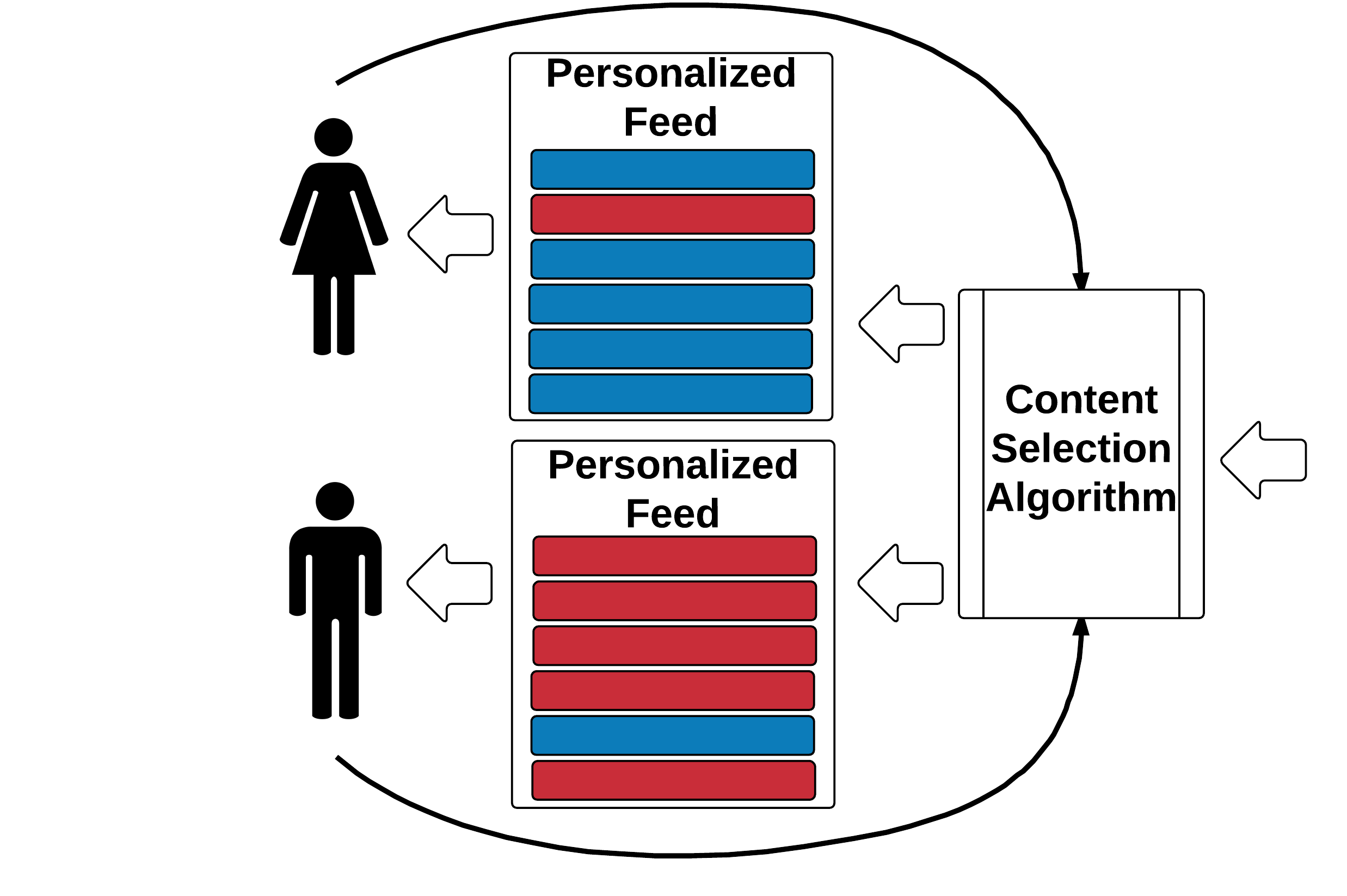}
		\caption{{Existing algorithms can perpetuate systemic bias by presenting different types of content to different types of users.}	\vspace{-.1in}}
		\label{fig:bias}
	\end{subfigure}
	\qquad 
	\begin{subfigure}[b]{0.45\textwidth}
	\centering
		\includegraphics[width=\textwidth]{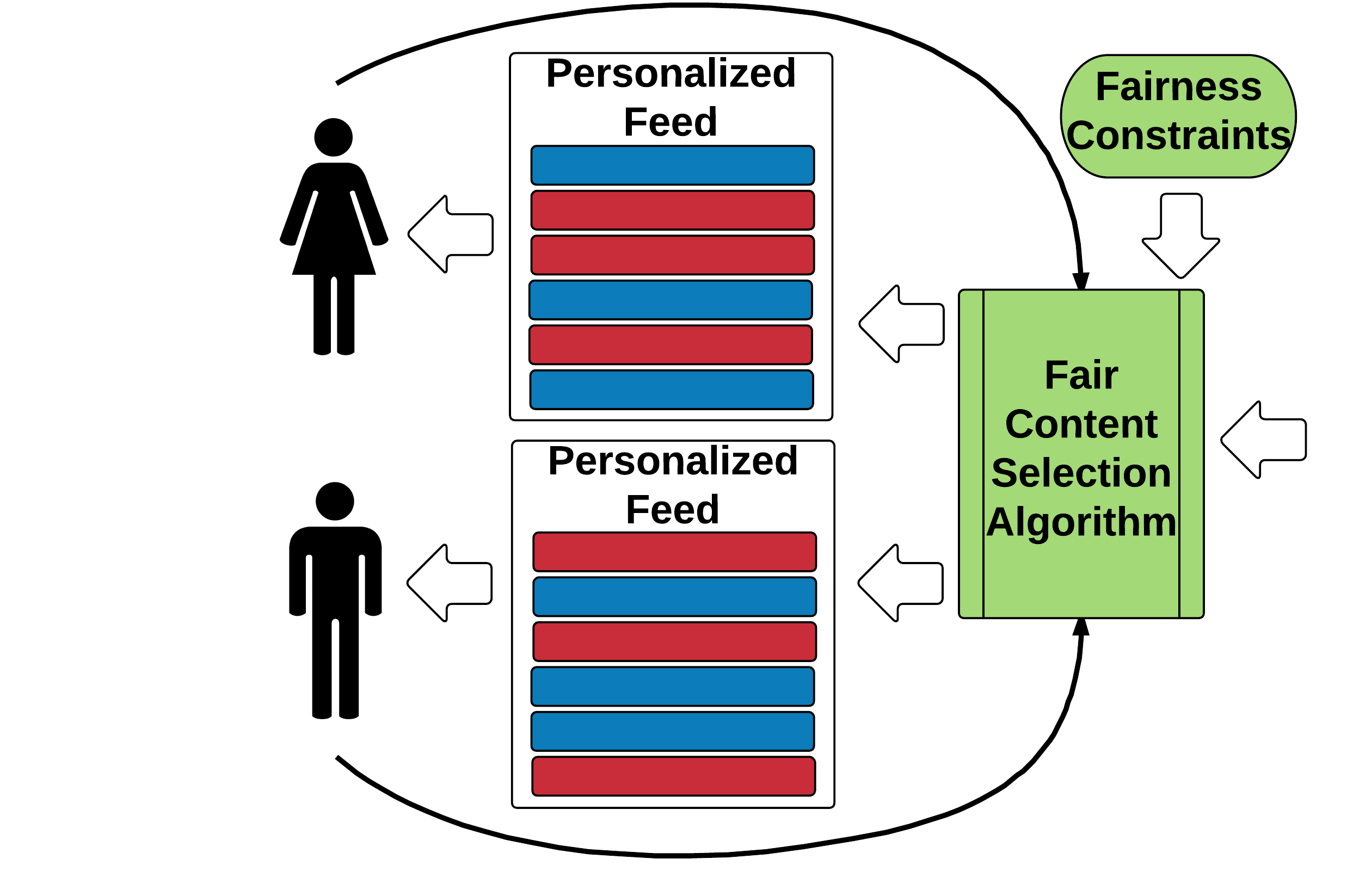}
		\caption{{Our proposed solution satisfies the fairness constraints and does not allow extreme differences in the types of content presented to different users.}	\vspace{-.1in}}
		\label{fig:fair}
	\end{subfigure}
		\caption{{Different colors in the content represent different groups, e.g., ads for high vs low paying jobs. Our fair content selection algorithm does not permit extreme biases while personalizing content.}}
	\label{fig:bias_fair}
	\vspace{-.1in}
\end{center}
\end{figure*}

One approach to eliminate such biases is to hide certain user properties so that they cannot be used for personalization; 
however, this could come at a significant loss to the utility for both the user and the platform -- as the content displayed would be less relevant on average and hence incur fewer clicks and result in decreased attention from the user and less revenue for the platform 
(see, e.g., \cite{sakulkar2016stochastic}). 
\emph{Can we design personalization algorithms that allow us to trade-off fairness with utility?}
We take a first step towards addressing this question in a rigorous algorithmic manner; for concreteness we describe our approach for personalized news feeds, however it also applies to other personalization settings.
Users have different {\em types} and content is classified into \emph{groups}; often the types can be defined using a small set of \emph{sensitive attributes} such as race, gender or age and the groups can similarly be defined by a small set of properties.
Current personalization algorithms, at every time-step, select a piece of content for the user,\footnote{In order to create a complete feed, content is simply selected repeatedly to fill the screen as the user scrolls down. The formalization does not change and hence, for clarity, we describe the process of selecting a single piece of content.} and feedback is obtained in the form of whether they click on, purchase or hover over the item.
The goal of the content selection algorithm is to select content for each user in order to maximize the positive feedback (and hence revenue) received.
As this optimal selection is a-priori unknown, the process {is often}  modeled as an online learning problem in which a probability distribution (from which one selects content) is maintained and updated according to the feedback \cite{pandey2006handling}.
As such content selection algorithms learn more about a user, the corresponding probability distributions become sparse (i.e., concentrates the mass on a small subset of entries); we propose that this is what leads to extreme personalization in which content feeds skew entirely to a single type of content.

To counter this, we introduce a notion of \emph{online group fairness}, in which we require that the probability distribution from which content is sampled  satisfies certain fairness constraints {\em at all time steps}; this in turn ensures that the probability vectors do not become sparse (or specialize) and the content shown to different types of users is similar with respect to the distribution of groups that it is drawn from.
We begin by considering a metric that aims to measure the fairness of a content selection algorithm. 
We can then define a set of constraints that, if satisfied by a content selection algorithm, guarantee a level of fairness according to this metric. 
We do not presume to define the constraints as setting them appropriately would depend on the application; instead, we take the constraints, types or sensitive attributes and groups as input (in the same spirit as \cite{Gummadi2015,FatML,yang2016measuring}). 
Subsequently we invoke the bandit convex optimization framework to solve the resulting algorithmic problem  of {\em maximizing revenue while satisfying fairness constraints}.
Our work gives rise to initial theoretical results which indicate that the structure of fairness constraints can be leveraged to obtain regret bounds that are not too far from the unconstrained setting.
Hence, this suggests that there is significant potential to de-bias personalization algorithms without a significant loss to utility.

\section{Our Contributions}
\subsection{Preliminaries}
Algorithms for the general problem of {displaying content} to users largely fall within the multi-armed bandit framework, and \emph{stochastic contextual bandits} in particular \cite{BanditBook}. 
At each time step $t = 1, \ldots, T$, a user views a {page} (e.g., Facebook, twitter or google news), its type (also known as its \emph{context}) $s^t \in \mathcal S$ is given as input, and one {piece of content} (or \emph{arm}) $a^t \in [k]$ must be selected to be displayed. 
A random \emph{reward} $r_{a^t, s^t}$ (hereafter denoted by $r_{a,s}^t$ for readability), which depends on both the given user type and the selected {content} is then received. 
{This reward captures resulting clicks, purchases, or time spent viewing the given content and depends not only on the type of user $s$ (e.g., men may be more likely to click on a sports article) but also on the content $a$ itself (e.g., some news articles have higher quality or appeal than others).}
More formally, at each time step $t$, a sample 
$$(s^t, r_{1,s}^t, \ldots, r_{k,s}^t)$$ is drawn from an unknown distribution $\mathcal D$, the context $s^t \in \mathcal S$ is revealed, the player selects an arm $a \in [k]$ and receives reward $r_{a,s}^t \in [0,1]$.
 As is standard in the literature, we assume that the $r_{a,s}$s are drawn independently across $a$ and $t$ (and not necessarily $s$).  
 The rewards $r_{a^\prime,s^\prime}$ for any $a^\prime \neq a$ and $s^\prime \neq s$ are assumed to be {\em unknown} -- indeed, there is no way to observe what a user's actions {would have been} had a different {piece of content} been displayed, or what a different user would have done.
An algorithm computes a probability distribution $p^{t}$ over the {arms} based on the previous observations $$(s^1, a^1, r_{a,s}^1), \ldots, (s^{t-1},a^{t-1},r_{a,s}^{t-1})$$ and the current user type $s^{t}$, and then selects {arm} $a^t \sim p^t$; as $p^t$ depends on the context $s^t$, we often write $p^t(s^t)$ for clarity. 
The goal is to select $p^t(s^t)$s in order to maximize the cumulative rewards, and 
the efficacy of such an algorithm is measured with respect to how well it minimizes \emph{regret} -- the difference between the algorithm's reward and the reward obtained from the (unknown) optimal policy.
Formally, let $f: \mathcal S \to [k]$ be a mapping from contexts to {arms}, and let $$f^\star := \arg\max_{f} \mathbb E_{(s,\vec{r})\sim\mathcal D} [r_{f(s),s}];$$ i.e., $f^\star$ is the policy that selects the best arm in expectation for each context. 
Then, the regret is defined as
$$  \mathsf{Regret} := T \cdot \mathbb E_{(s,\vec{r})\sim\mathcal D} [r_{f^\star(s),s}] - \sum_{t=1}^T r_{a^t,s^t}. $$
Note that the regret is a random variable as $a^t$ depends not only the draws from $p^t$, but also on the realized \emph{history} of samples $\mathcal H^T = \{(s^t, a^t, r_{a,s}^t)\}_{t=1}^T$.

\subsection{\bf Group Fairness}
Towards defining group fairness in the setting above, 
let $G_1,$ $\ldots,$ $G_g \subseteq [k]$ be $g$ \emph{groups} of {content}. 
{For instance the $G_i$s could form a partition (e.g., ``republican-leaning'' news articles, ``democratic-leaning'', and ``neutral'').
An important feature of bandit algorithms, which ends up being problematic for fairness,  is that the probability distribution converges to the action with the best expected reward for each context; i.e., the entire probability mass in each context ends up on a single group. 
Thus different types of users may be shown very different ad groups (Figure~\ref{fig:bias}), and, e.g., only show minimum-wage jobs to disenfranchised populations.} 

\subsubsection*{\bf A fairness metric} Given an algorithm for the problem as above, we consider the following metric for its fairness.
Let
\begin{equation}\label{eq:def}
\alpha_i^t := 1 - \sup_{\mathcal H^{t-1}} \left\{ \max_{s_1, s_2 \in \mathcal S} \left(\sum_{a \in G_i} p^t_a(s_1) -  \sum_{a \in G_i} p^t_a(s_2)\right) \right\},
\end{equation}
for all $i \in [g]$ and $t \in [T]$, where $p^t_a(s)$ is the probability with which the algorithm selects {arm} $a$ at time $t$ given the context $s$ and $\mathcal H^{t-1}$ is the realized (stochastic) history up to time $t-1$.
Note that $\alpha_i^t \in [0,1]$ and measures how fair the algorithm is to group $i$ until time $t$, with $\alpha_i^t=0$ representing extreme unfairness and $\alpha_i^t=1$ representing perfect fairness.
The supremum over all possible stochastic trajectories (histories) ensures that this metric measures the worst case discrepancy in the probability mass assigned to {the arms} in group $i$ for every pair of distinct contexts. 
Given  definition \eqref{eq:def}, we define the fairness with respect to group $i$ as 
\begin{equation}\label{eq:2} \alpha_i := \min_{t \in [T]} \alpha_i^t,
\end{equation}
 and the group fairness of the algorithm as $$\alpha := \min_{i \in [k]} \alpha_i.$$ 
Depending up on how stringent one would like to be, the supremum in \eqref{eq:def} could be relaxed to an expectation over the history; however,
 the latter would allow specific instances of the algorithm to select probability distributions that are highly biased.
One could also consider variants that use different measures of discrepancy across contexts of the probability mass assigned to a group  as opposed to the difference as in \eqref{eq:def}.

\subsubsection*{\bf Our fairness constraints.}
Motivated by the metrics \eqref{eq:def} and \eqref{eq:2}, we define our \emph{group fairness constraints} as follows  
\begin{equation}
\label{eq:fair}
   \ell_i \leq \sum_{a \in G_i} p^t_a(s) \leq   u_i \;\;\;\; \forall i \in [g], \forall t \in [T], \forall s \in \mathcal S.
\end{equation}
In other words, we ensure that the probability mass placed on any given group is neither too high nor too low at each time step.
Clearly, an algorithm that satisfies the fairness constraints has 
$$\alpha_i \geq 1-(u_i -   \ell_i)$$ fairness with respect to group $i$ by definition. 
As discussed above, rather than specifying the values of $u_i$s and $\ell_i$s ourselves, we allow them to be specified as input as appropriate values may vary significantly depending on the application or desired result.
Having the $u_i$s and $\ell_i$s as part of the input to the algorithm allows one to control the extent of group fairness depending on the application.
Unlike ignoring user types entirely, our constraints still allow for personalization \emph{across} groups. 
For instance, if the groups are republican (R) vs democrat (D) articles, and the user types are known republicans (r) or democrats (d), we may require that $p^t_{\mbox{R}}(\cdot) \leq 0.75$ and $p^t_{\mbox{D}}(\cdot) \leq 0.75$ for all $t$. 
This ensures that extreme polarization cannot occur -- at least 25\% of the articles a republican is presented with will be democrat-leaning.
Despite these constraint, personalization at the group level can still occur, e.g., by letting $p_{\mbox{R}}^t(\mbox{r}) = 0.75$ and $p_{\mbox{R}}^t(\mbox{d}) = 0.25$.
Furthermore, this allows for complete personalization \emph{within} a group -- the republican-leaning articles shown to republicans and democrats may differ; this is crucial for our setting as the utility maximizing articles for republicans vs democrats, within a group, may differ.

To the best of our knowledge, this notion of fairness is novel and addresses the concerns in the motivating examples.
 In a different context one may seek a different notion of fairness.
For instance, if the options are people rather than news stories, one would want to be fair to each individual and such notions of \emph{online individual fairness} have recently been developed \cite{joseph2016fairness}. 
Thus, it is the particulars of the online personalization of content that motivates this definition of group fairness. 

\vspace{-.05in}
\subsection{Algorithmic Results}
\subsubsection*{\bf Group fair regret}
We measure an algorithm's performance against the best \emph{fair} solution.\footnote{The unconstrained regret may be arbitrarily bad, e.g., if $u_i = \vareps \ll 1$ for the arm with the best reward.} 
We say that a distribution $p$ is \emph{fair} if it satisfies the upper and lower bound constraints in \eqref{eq:fair}, and let $\mathcal C$ be the set of all such distributions.  
Let $\mathcal B$ be the set of functions $g: \mathcal S \to [0,1]^k$ such that $g(s) \in \mathcal C$; i.e., all $g \in \mathcal B$ satisfy the fairness constraints.  
Further, we let 
$$g^\star := \arg\max_{g \in \mathcal B} \mathbb E_{(s,\vec{r})\sim\mathcal D} [r_{g(s),s}];$$  i.e., $g^\star$ is the policy that selects the best arm in expectation for each context. 
An algorithm is said to be fair if it only selects $p^t(s^t) \in \mathcal C$.
Thus, the \emph{fair regret} for such an algorithm can be defined as
\[  \mathsf{FairRegret} := T \cdot \mathbb E_{(s,\vec{r})\sim\mathcal D} [r_{g^\star(s),s}] - \sum_{t=1}^T r_{a^t,s^t}. \]

 \vspace{-.05in}
\subsubsection*{\bf Regret bounds.} %
We show how techniques developed in the context of stochastic bandit optimization, along with the special structure of the constraint set $\mathcal{C}$, can yield a (roughly) ${\sf poly} \log T$ regret guarantee via an efficient algorithm.
For each arm $a \in [k]$ and each context $s \in \mathcal{S}$, let its mean reward be $\mu_{a,s}$. 
We first consider the case of a single context ($|\mathcal S| = 1$) and then briefly discuss how this can be extended to the general case using standard techniques. 
In this case, the unknown parameters are the expectations of each arm $\mu_a$ for $a \in [k]$.
In particular, we rely on an algorithm and  analysis developed in  \cite{dani2008stochastic}; restated in the following theorem.
\vspace{-.05in}
\begin{theorem}
There is a polynomial time algorithm that, given the description of $\mathcal{C}$ and the sequence of rewards, obtains the following regret bound:
$$\mathbb{E}\left[{\sf FairRegret}\right]=O\left(\frac{k^3\log^3 T}{\gamma}\right),$$ 
where the expectation is taken over the histories and $a^t \sim p^t$, and $\gamma$ depends on $(\mu_{a})_{a \in [k]}$ and $\C$ as defined in \eqref{eq:gamma}.
\label{thm:abcd1}
\end{theorem} 
\vspace{-.05in}
\noindent
The quantity $\gamma$ is the difference between the maximum and the second maximum of the expected reward with respect to the $\mu$s over the vertices of the polytope $\mathcal{C}$.
Formally, let $V(\mathcal{C})$ denote the set of vertices of $\mathcal{C}$ and 
$$v^\star := \arg\max_{v \in V(\mathcal{C})} \sum_{a \in [k]} \mu_a v_a.$$ 
Then,
\begin{equation}\label{eq:gamma}
\gamma:=  \sum_{a \in [k]} \mu_a v_a^\star - \max_{v \in V(\mathcal{C}) \backslash v^\star} \sum_{a \in [k]} \mu_a v_a. 
\end{equation}
For general convex sets, $\gamma$ can be $0$ and the regret bound can at best only be $\sqrt{T}$ \cite{dani2008stochastic}.
The fact that our fairness constraints imply that $\mathcal{C}$ is a polytope implies that, unless there are degeneracies, $\gamma$ is non-zero.

Further, we can quickly obtain quantitative bounds on $\gamma$ in our setting assuming some structure on the groups.
Here, we present a rough sketch of these results.

To start with, let $$T_a:=\{i \in [g] : a \in G_i \}$$ be the set of properties that {arm} $a$ has and let  $$D := \max_{a \in [k]} |T_a|.$$
In the simplest case of practical interest, $D=1$, i.e., the groups of {arms} are {\em disjoint}.
Here, we can argue that the constraints defining a $v \in V(\mathcal{C})$ can be chosen so that the corresponding matrix of equalities that determines it is totally unimodular.
Let $\mu_a=M_a/M$ for some positive integers $M_a,M$ and assume that $M_a \neq M_{a'}$ for all $a \neq a' \in [k]$.
Similarly, let $\ell_i=L_i/N$ an $u_i=U_i/N$ be their rational representation.
By randomly perturbing $L_i$s and $U_i$s we can ensure that $v^\star$ is unique with high probability; implying that $\gamma \neq 0$.
The unimodularity property for $D=1$ implies that 
$ \gamma \geq \frac{1}{MN}.$
This is because a vertex $v=A^{-1}b$ where $A$ is a totally unimodular matrix and $b$ is a vector consisting of $\ell_i$s, $u_i$s, $1$ or $0$. 
Hence, if the numbers $\ell_i$ and $u_i$ are not too small or too large, our fairness constraints result in a sufficiently large $\gamma$, resulting in low regret.
Note that even in the unconstrained case, we cannot hope that $\gamma \geq 1/M$ as the difference between the expected rewards of the best two arms could be that small. 
When $D>1,$ the constraint matrices corresponding to vertices may no longer totally unimodular. 
However, as each arm appears in at most $D$ groups, we can upper bound its determinant by roughly $D^g$; this gives a lower bound of $D^g$ in the denominator of the coordinates of any vertex.
Consequently, one can extend the above argument to prove that if $\gamma$ is nonzero, then  
$\gamma \geq  \frac{1}{D^g} \frac{1}{MN}.\label{eq:fsa}$

The result in Theorem \ref{thm:abcd1} can be easily extended to the case of $|\mathcal S| > 1$ by running, for instance, the algorithm for each context separately; see Theorem 4.1 in \cite{BanditBook}.
The regret bounds worsen by $|\mathcal S|$. 
As we anticipate $|\mathcal S|$ to be a small as we only need to be fair with respect to sensitive attributes (e.g., gender or race) as opposed to all contexts. %

The definitions of fairness and the constraints can also be applied without modification to the adversarial regime. The constrained problem then falls in the regime of bandit convex optimization \cite{hazan2016introduction} -- we omit the details.

\section{Related Work}
There have been studies proposing notions of group fairness such as {statistical parity} and {disparate impact} \cite{kamiran2009classifying, Gummadi2015, feldman2015certifying}, they apply to the \emph{offline} problem; in our setting this would correspond to enforcing $p^T(s)$ to be roughly the same for all $s$, but would leave the intermediary $p^t(s)$ for $t < T$ unrestricted. 
A subtle point is that most notions of (offline) group fairness primarily consider the selection or classification of groups of \emph{users}; in our context, while the goal is still fairness towards users, the selection is over \emph{content}. However, as the  constraints necessary to attain fairness remain on the selection process, we use the terminology  \emph{online group fairness} to highlight these parallels.

A recent work \cite{joseph2016fairness} defined a notion of \emph{online individual fairness}  which restricts $p^t$s so that all \emph{arms} are treated equally by only allowing the probability of one arm to be more than another if we are reasonably certain that it is better than the other.
When the arms correspond to \emph{users} and not content such individual fairness is indeed crucial, but for the personalized setting the requirement is both too strong (we are only concerned with \emph{groups} of content) and too weak (it still allows for convergence to different groups for different contexts). 

Other constrained bandit settings that encode \emph{global} knapsack-like constraints (and locally place no restriction on $p^t$) have also been considered; see, e.g., \cite{agrawal2016linear}.

\section{Discussion \& Future Work}
In summary, motivated by personalization, we initiate the study of group fairness in online settings. 
We propose a definition for group fairness, and define constraints for online algorithms that ensure fairness guarantees.
Our constraints suffice to ensure group fairness but may not be necessary; exploring this remains an important open question.
Further, we present an algorithmic framework for group fairness in online learning by appealing to the contextual bandit optimization framework. 
We show that well-known results from the bandit convex optimization literature can be used along with the structure of the fairness constraint set in order to obtain poly-logarithmic regret in the stochastic bandit setting. 
On a theoretical front, it would be an important avenue of future work to develop specialized algorithms for such constraints that optimize both the regret and the running time.

This work applies to many personalization settings. For instance, in online advertising, ad agencies can select what type of user profiles they desire an ad to be shown to. 
Additionally, the ad exchange (i.e., the platform that matches ads to users) further learns about which ads different types of users are more likely to click on and uses this information in its decision-making process (see \cite{muthukrishnan2009ad} for an overview). 
This personalization leads to higher utility, efficiency, and hence revenue \cite{farahat2012effective}, but has similar problems with regard to biases; recent work has shown that the gender of a user affects whether ads for high- vs. low-paying jobs are displayed \cite{datta2015automated, sweeney2013discrimination}.
Our formalization and approach applies directly to the  online advertising setting. 

One implicit assumption was that the groups of content and types of users were given. 
In most personalization settings we would expect the types of users to be known -- in particular, the sensitive attributes (such as gender, race, income, etc.) would already be encoded as features. More generally one could cluster the feature space into types depending on the application.
Content, however, may not be already clustered. While in some settings, such as online ads, we may already have the necessary features as they are used as part of the advertising exchange, in others settings, such as news feeds, this may not be the case.
 Here one practical challenge would be to additionally set up the necessary classification tools in order to group content appropriately.

Another key question is which constraints to apply to which groups. 
In order to ensure \emph{exact} fairness, all types of users see the same fraction $c_i$ of content group $i$. This can be achieved by letting $\ell_i = u_i = c_i$ for all $i$.
In such cases, personalization can occur \emph{within} groups of content, but not \emph{accross}. 
While this may be feasible in some settings, in others this may come at a significant loss to revenue and hence a balance between fairness and utility may have to be struck; what this balance looks like is an important question to study for individual applications.
Furthermore, the question of how to determine the optimal constraints $c_i$ (or $\ell_i, u_i$ more generally) is not obvious. 

The most   important open direction is to understand how such fair algorithms behave in practice. 
Given a set of constraints, how fair are the resulting algorithms (according to the fairness metric) for natural reward distributions? 
Perhaps more importantly, how much does fair personalized content affect user's perceptions and decisions, in particular as a function of the imposed constraints? 
Setting up such a field study would be crucial in order to determine the efficacy of this approach.


\bibliographystyle{plain}
\bibliography{references}

\end{document}